# Plane-Wave Reflection and Transmission at Arbitrarily Oriented and Charged Planar Interfaces between Lossy Isotropic Media

Zhili Lin (林志立), *Senior Member, IEEE*

***Abstract*—A general algorithm for calculating the reflection and refraction of nonuniform plane waves from an arbitrarily oriented and charged planar interface between two lossy isotropic media is proposed based on the decomposition of the complex wave vector and complex wave numbers with respect to the unit normal vector of the interface. According to the complex vector analysis, the exact definition of the complex angles of incidence, reflection and refraction are presented and applied in the complex forms of Snell' s law and Fresnel equations to quickly and correctly calculate the complex wave vectors and the complex electric fields of the reflected and refracted waves at a charged interface where the surface charge densities are considered. Two practical examples are given to demonstrate the validity and effectiveness of the proposed methodology. The method may find applications in the fields of remote sensing, light ray-tracing and the interaction of electromagnetic waves with absorbing media.**

***Index Terms*—Nonuniform plane wave, complex wave vector, charged interface, Snell's law, Fresnel equations, absorbing media**

## I. Introduction

THE reflection and refraction phenomena of nonuniform electromagnetic plane waves incident on planar interfaces between isotropic and possibly lossy media are a widely studied topic in electromagnetics and optics [1], which are applied in a variety of areas such as leaky-wave antenna [2], ray tracing [3]-[5], interaction with biological tissues [6], [7] and metasurfaces [8]. Nonuniform plane waves exist in various situations, such as the evanescent wave in total reflection at the interface of two lossless media when the incident angle is larger than the critical angle, the transmitted wave in the case that a uniform plane wave impinges upon an interface between a lossless medium and a lossy medium, or the incident, reflected and transmitted waves in the most complicated that a nonuniform plane wave impinges upon a planar interface between two lossy medium. As a practical example, the nonuniform wave can be produced by the transmission of a uniform wave incident on a prism made of lossy material [9].

The complex wave vectors of nonuniform plane waves and the complex angles in the complex forms of Snell's law and Fresnel equations are difficult to understand because they don't have physical meanings. The representation of a nonuniform plane wave by a complex wave vector $\boldsymbol{k}$ split up into two vectors called the phase vector $\boldsymbol{\beta}$ and the attenuation vector $\boldsymbol{\alpha}$ was introduced in details by Adler, Chu and Fano [10]. The phase vector $\boldsymbol{\beta}$ and the attenuation vector $\boldsymbol{\alpha}$ are real-valued and have physically meaningful directions. The problem of a planar interface between two lossy half spaces has been treated with the Adler–Chu–Fano formulation by Holmes [11] for the case where the incident plane wave was uniform, and by Radcliff [12] for the more general case where the incidence plane wave was nonuniform but $\boldsymbol{\beta}$, $\boldsymbol{\alpha}$ and the unit normal vector of the interface $\boldsymbol{e}_{\mathrm{n}}$ were coplanar. The extension to the more challenging case when the three vectors, $\boldsymbol{\beta}$, $\boldsymbol{\alpha}$ and $\boldsymbol{e}_{\mathrm{n}}$, are not coplanar is performed by Roy [13][14]. An easy analytic description to calculate the nonuniform wave vector from the phase-attenuation vector notation by employing real polarization vectors was introduced by Frezza and Tedeschi [15] and the ambiguity problem in the calculation of the transmission angles of the phase-and attenuation direction vectors are resolved by considering the direction of power flow propagation through the interface. Following this Mangini and Frezza applied those results to propose a transfer matrix-based method for calculation of the reflection and transmission coefficients at the interfaces of a stratified lossy medium [16]. Meanwhile, the deeply penetrating phenomena in lossy media without attenuation away from the interface and the critical angles for the deep penetration conditions are also studied with numerous applications [17]-[20].

Another popular notation of the nonuniform wave vector in dependence of a complex angle has been presented by Jackson [21]. The complex angle formulation of the nonuniform wave vector presented by Jackson, has been used by Dupertius, Proctor and Acklin to describe the polarization property of a nonuniform plane wave and to extend Snell's law and Fresnel equations for the case of noncoplanar phase and attenuation vectors [22]. The generalized laws of reflection and refraction for nonuniform plane waves at interfaces of lossy isotropic media are presented by Fedorov and Nakajima to study the negative refraction of nonuniform plane waves at the properly

Manuscript received March 10, 2025. This work was supported in part by the Natural Science Foundation of Xiamen City of China under Grant 3502Z202473050 and by the National Natural Science Foundation of China under Grant 61101007.

The author is with the Fujian Key Laboratory of Light Propagation and Transformation, College of Information Science and Engineering, Huaqiao University, Xiamen 361021, China (e-mail: zllin2008@gmail.com).

oriented interfaces of conventional materials [23]. The equality of Fresnel coefficients derived by the complex angle representation and those derived by the phase- and attenuation vector representation was doubted by Canning [24] and finally demonstrated by Besieris [25]. Employing the complex angle notation of the nonuniform wave vector, Weber introduced an effective transmission coefficient to adapt the formula of energy conservation at an interface to lossy systems when $\boldsymbol{\beta}$, $\boldsymbol{\alpha}$ and $\boldsymbol{e}_{\mathrm{n}}$ are coplanar [26]. An overview concerning the publications, methods and notations used to describe the effective propagation constants of non-uniform plane waves at a lossy interface is presented by Schake [27]. Moreover, the external surface charges on the interface also have a significant impact on the reflection and refraction properties of incident electromagnetic waves. Based on the concept of excess charge surface conductivity, the generalized laws of Snell, Fresnel and energy balance are derived by Zhang *et al.* [28] for the harmonic nonuniform plane waves that are incident upon a charged plane interface between two absorbing media.

In this work, the fast method for calculating the reflected and transmitted waves is presented for a given nonuniform plane wave incident on an arbitrarily oriented and charged planar interface between two isotropic and possibly lossy media. The novelty of this work lies in the decomposition of the complex wave vectors and complex wave numbers of the nonuniform plane waves with respect to the interface plane, which greatly simplifies the calculation of the complex angles and complex wave vectors of the reflected and transmitted nonuniform plane waves without the need to solve the various equations about the real direction angles of $\boldsymbol{\beta}$ and $\boldsymbol{\alpha}$. Based on the proposed decomposition and definition, the Snell's law and Fresnel equations can be conveniently and correctly applied to calculate the reflected and transmitted waves.

The subsequent parts of this work are organized as follows. In Section II, the decomposition of the complex wave vector and complex wave number of a nonuniform plane wave with respect to an arbitrarily oriented plane is performed along the normal direction and the tangential direction based on the complex vector analysis. Meanwhile, the complex angle of the complex wave vector with respect to the unit normal vector of an arbitrarily oriented plane is defined firstly in literature. In Section III, the relationship of the complex wave vector and the propagation constants of a nonuniform plane wave to the intrinsic material parameters are presented and the validity of the above decomposition is demonstrated. In Section IV, the complex electric fields are formulated based on the fact that an arbitrary nonuniform plane wave can be considered as the combination of two polarized waves, the so-called parallel electric (PE) wave and the parallel magnetic (PM) wave. The decomposition of the complex wave vector and complex wave number is applied to the Snell's law and therefore the complex angles and the complex wave vectors of the incident, reflected and transmitted waves can be easily and quickly calculated. In Section V, Fresnel equations for calculating the reflection and transmission coefficients of the PE polarized and PM polarized incident plane waves at a charged planar interface are given in an extremely concise form including the surface conductivity. In Section VI, the expressions for the time-averaged complex Pointing vectors are presented and the energy balance relation is given. In Section VII, the calculation procedure for the reflected and transmitted waves is presented for a given nonuniform plane wave incident on an arbitrarily oriented and charged planar interface between two isotropic lossy media. In Section VIII, a two-dimensional (2D) example of a charged prism and a three-dimensional (3D) example of a cuboid composed of two lossy and charged blocks are provided to demonstrate the simplicity and powerfulness of our proposed method. In the last Section, the conclusion is made to this work.

## II. Complex Wave Vector Decomposition

Suppose that a monochromatic nonuniform electromagnetic plane wave with the time dependence $\mathrm{e}^{-\mathrm{j}\omega t}$ is propagating in a homogeneous, isotropic and linear medium. For the convenience of understanding, the complex wave vector of the nonuniform plane wave is often expressed as the superposition of two real-valued vectors with physically meaningful directions,

$$\boldsymbol{k} = \boldsymbol{\beta} + \mathrm{j}\boldsymbol{\alpha} \quad (1)$$

where $\boldsymbol{\beta}$ is the phase vector perpendicular to the planes of constant phase and $\boldsymbol{\alpha}$ is the attenuation vector perpendicular to the planes of constant magnitude. For a nonuniform plane wave, $\boldsymbol{\beta}$ and $\boldsymbol{\alpha}$ are not parallel to each other and $\boldsymbol{k}$ doesn't have a physically meaningful direction. However, for a uniform plane wave, $\boldsymbol{\beta}$, $\boldsymbol{\alpha}$ and $\boldsymbol{k}$ are all parallel and share the same direction.

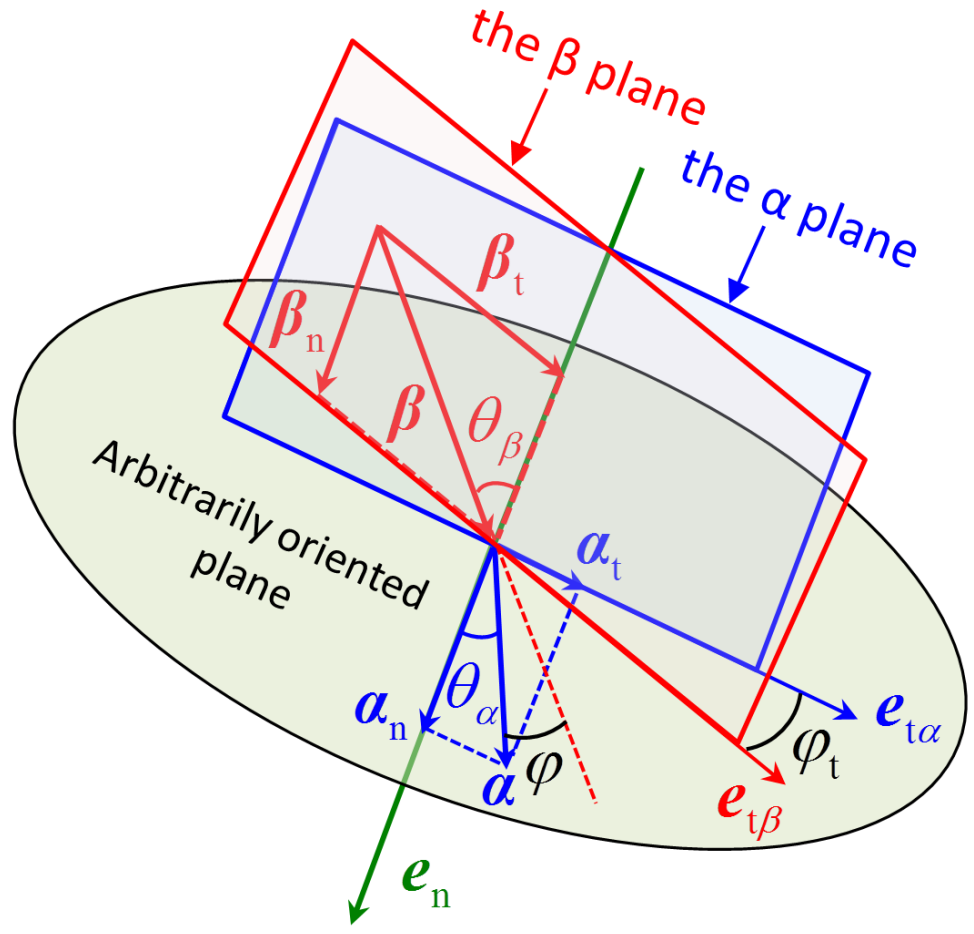

Fig. 1. The decomposition of the phase vector $\boldsymbol{\beta}$ and the attenuation vector $\boldsymbol{\alpha}$ of a nonuniform plane wave along the normal and tangential directions of an arbitrarily oriented plane for the most challenging case where $\boldsymbol{\beta}$, $\boldsymbol{\alpha}$ and the unit normal vector $\boldsymbol{e}_{\mathrm{n}}$ are not coplanar. The $\beta$ plane is formed by $\boldsymbol{\beta}$ and $\boldsymbol{e}_{\mathrm{n}}$, while $\theta_\beta$ is the angle between $\boldsymbol{\beta}$ and $\boldsymbol{e}_{\mathrm{n}}$. The $\alpha$ plane is formed by $\boldsymbol{\alpha}$ and $\boldsymbol{e}_{\mathrm{n}}$, while $\theta_\alpha$ is the angle between $\boldsymbol{\alpha}$ and $\boldsymbol{e}_{\mathrm{n}}$. $\varphi$ is the angle between $\boldsymbol{\alpha}$ and $\boldsymbol{\beta}$, while $\varphi_{\mathrm{t}}$ is the angle between the tangent component vector $\boldsymbol{\alpha}_{\mathrm{t}}$ of $\boldsymbol{\alpha}$ and the tangent component vector $\boldsymbol{\beta}_{\mathrm{t}}$ of $\boldsymbol{\beta}$.

As shown in Fig. 1, an arbitrarily oriented plane of interface has a unit normal vector $\boldsymbol{e}_\mathrm{n}$ is assumed. Then the phase vector $\boldsymbol{\beta}$ can be decomposed with respect to the $\beta$ plane into two component vectors along the normal direction $\boldsymbol{e}_\mathrm{n}$ and the tangential direction $\boldsymbol{e}_{\mathrm{t}\beta}$,

$$\boldsymbol{\beta} = \boldsymbol{\beta}_\mathrm{n} + \boldsymbol{\beta}_\mathrm{t} = \beta_\mathrm{n}\boldsymbol{e}_\mathrm{n} + \beta_\mathrm{t}\boldsymbol{e}_{\mathrm{t}\beta} \tag{2}$$

where $\boldsymbol{\beta}_\mathrm{n} = (\boldsymbol{e}_\mathrm{n}\cdot\boldsymbol{\beta})\boldsymbol{e}_\mathrm{n}$, $\boldsymbol{\beta}_\mathrm{t} = \boldsymbol{\beta} - \boldsymbol{\beta}_\mathrm{n}$ and $\theta_\beta = \arccos(\beta_\mathrm{n}/\beta)$. Similarly, the attenuation vector $\boldsymbol{\alpha}$ also can be decomposed into the form

$$\boldsymbol{\alpha} = \boldsymbol{\alpha}_\mathrm{n} + \boldsymbol{\alpha}_\mathrm{t} = \alpha_\mathrm{n}\boldsymbol{e}_\mathrm{n} + \alpha_\mathrm{t}\boldsymbol{e}_{\mathrm{t}\alpha} \tag{3}$$

where $\boldsymbol{\alpha}_\mathrm{n} = (\boldsymbol{e}_\mathrm{n}\cdot\boldsymbol{\alpha})\boldsymbol{e}_\mathrm{n}$, $\boldsymbol{\alpha}_\mathrm{t} = \boldsymbol{\alpha} - \boldsymbol{\alpha}_\mathrm{n}$ and $\theta_\alpha = \arccos(\alpha_\mathrm{n}/\alpha)$. When the three vectors $\boldsymbol{\alpha}$, $\boldsymbol{\beta}$ and $\boldsymbol{e}_\mathrm{n}$ are coplanar, $\boldsymbol{e}_{\mathrm{t}\beta} = \boldsymbol{e}_{\mathrm{t}\alpha}$. If $\boldsymbol{\alpha}$, $\boldsymbol{\beta}$ and $\boldsymbol{e}_\mathrm{n}$ are not coplanar, $\boldsymbol{e}_{\mathrm{t}\beta} \neq \boldsymbol{e}_{\mathrm{t}\alpha}$. The latter scenario is depicted in Fig. 1 and it is also the general case for a nonuniform plane wave incident on an arbitrarily oriented interface between two different media.

Now consider the decomposition of the complex wave vector with respect to an arbitrarily oriented plane with a unit normal vector $\boldsymbol{e}_\mathrm{n}$ by using the complex vector identity,

$$\boldsymbol{k} = (\boldsymbol{e}_\mathrm{n}\cdot\boldsymbol{k})\boldsymbol{e}_\mathrm{n} + (\boldsymbol{e}_\mathrm{n}\times\boldsymbol{k})\times\boldsymbol{e}_\mathrm{n} = \boldsymbol{k}_\mathrm{n} + \boldsymbol{k}_\mathrm{t} \tag{4}$$

where $\boldsymbol{k}_\mathrm{n} = (\boldsymbol{e}_\mathrm{n}\cdot\boldsymbol{k})\boldsymbol{e}_\mathrm{n}$ is the component vector normal to the plane and $\boldsymbol{k}_\mathrm{t} = (\boldsymbol{e}_\mathrm{n}\times\boldsymbol{k})\times\boldsymbol{e}_\mathrm{n} = \boldsymbol{k} - \boldsymbol{k}_\mathrm{n}$ is the component vector parallel to the plane. To under the mathematical meanings of $\boldsymbol{k}_\mathrm{n}$ and $\boldsymbol{k}_\mathrm{t}$, we get from (1), (2) and (3) that

$$\boldsymbol{k} = \boldsymbol{\beta} + \mathrm{j}\boldsymbol{\alpha} = (\boldsymbol{\beta}_\mathrm{n} + \mathrm{j}\boldsymbol{\alpha}_\mathrm{n}) + (\boldsymbol{\beta}_\mathrm{t} + \mathrm{j}\boldsymbol{\alpha}_\mathrm{t}) = \boldsymbol{k}_\mathrm{n} + \boldsymbol{k}_\mathrm{t} \tag{5}$$

where

$$\boldsymbol{k}_\mathrm{n} = (\boldsymbol{e}_\mathrm{n}\cdot\boldsymbol{k})\boldsymbol{e}_\mathrm{n} = \boldsymbol{\beta}_\mathrm{n} + \mathrm{j}\boldsymbol{\alpha}_\mathrm{n} = (\beta_\mathrm{n} + \mathrm{j}\alpha_\mathrm{n})\boldsymbol{e}_\mathrm{n} = k_\mathrm{n}\boldsymbol{e}_\mathrm{n} \tag{6}$$

and

$$\boldsymbol{k}_\mathrm{t} = \boldsymbol{k} - \boldsymbol{k}_\mathrm{n} = \boldsymbol{\beta}_\mathrm{t} + \mathrm{j}\boldsymbol{\alpha}_\mathrm{t} = \beta_\mathrm{t}\boldsymbol{e}_{\mathrm{t}\beta} + \mathrm{j}\alpha_\mathrm{t}\boldsymbol{e}_{\mathrm{t}\alpha} = k_\mathrm{t}\boldsymbol{e}_\mathrm{t} \tag{7}$$

It is noted that $\boldsymbol{e}_\mathrm{n}$ is the real unit vector of $\boldsymbol{k}_\mathrm{n}$ with a physically meaningful direction normal to the plane. However, $\boldsymbol{e}_\mathrm{t}$ is the complex unit vector of $\boldsymbol{k}_\mathrm{t}$ composed of $\boldsymbol{e}_{\mathrm{t}\beta}$ and $\boldsymbol{e}_{\mathrm{t}\alpha}$ in a complex form, which satisfies $\boldsymbol{e}_\mathrm{t}\cdot\boldsymbol{e}_\mathrm{t} = 1$ but hasn't a physically meaningful direction. Since $\boldsymbol{e}_\mathrm{n} \perp \boldsymbol{e}_{\mathrm{t}\beta}$ and $\boldsymbol{e}_\mathrm{n} \perp \boldsymbol{e}_{\mathrm{t}\alpha}$ from Fig. 1, one obtains $\boldsymbol{e}_\mathrm{n}\cdot\boldsymbol{e}_\mathrm{t} = 0$ and $\boldsymbol{k}_\mathrm{n}\cdot\boldsymbol{k}_\mathrm{t} = 0$.

Assume that $\boldsymbol{k} = k\boldsymbol{e}_k$, where $\boldsymbol{e}_k$ is the complex unit vector of $\boldsymbol{k}$ satisfying $\boldsymbol{e}_k\cdot\boldsymbol{e}_k = 1$, but it is also without a physically meaningful direction like that of $\boldsymbol{e}_\mathrm{t}$. Then from (5) we get

$$k^2 = \boldsymbol{k}\cdot\boldsymbol{k} = (\boldsymbol{k}_\mathrm{n} + \boldsymbol{k}_\mathrm{t})\cdot(\boldsymbol{k}_\mathrm{n} + \boldsymbol{k}_\mathrm{t}) = \boldsymbol{k}_\mathrm{n}\cdot\boldsymbol{k}_\mathrm{n} + \boldsymbol{k}_\mathrm{t}\cdot\boldsymbol{k}_\mathrm{t} = k_\mathrm{n}^2 + k_\mathrm{t}^2 \tag{8}$$

where $k_\mathrm{t}$ is the tangential component of wave number given by

$$k_\mathrm{t} = \sqrt{\boldsymbol{k}_\mathrm{t}\cdot\boldsymbol{k}_\mathrm{t}} = \sqrt{(\boldsymbol{\beta}_\mathrm{t} + \mathrm{j}\boldsymbol{\alpha}_\mathrm{t})\cdot(\boldsymbol{\beta}_\mathrm{t} + \mathrm{j}\boldsymbol{\alpha}_\mathrm{t})} \tag{9}$$

Since $\cos^2 z + \sin^2 z = 1$ still holds for a complex variable $z$, the complex angle $\theta$ of $\boldsymbol{k}$ with respect to $\boldsymbol{e}_\mathrm{n}$ is defined as

$$\theta = \arcsin(k_\mathrm{t}/k) \tag{10}$$

and we have

$$k^2 = k_\mathrm{n}^2 + k_\mathrm{t}^2 = k^2\cos^2\theta + k^2\sin^2\theta \tag{11}$$

where the normal component $k_\mathrm{n}$ is obtained by

$$k_\mathrm{n} = k\cos\theta \tag{12}$$

while the tangential component $k_\mathrm{t}$ is obtained by

$$k_\mathrm{t} = k\sin\theta = \sqrt{\boldsymbol{k}_\mathrm{t}\cdot\boldsymbol{k}_\mathrm{t}}. \tag{13}$$

It is noted that $\boldsymbol{k}_\mathrm{t}$ can be calculated by $\boldsymbol{k}_\mathrm{t} = \boldsymbol{k} - (\boldsymbol{e}_\mathrm{n}\cdot\boldsymbol{k})\boldsymbol{e}_\mathrm{n}$ or by $\boldsymbol{k}_\mathrm{t} = (\boldsymbol{e}_\mathrm{n}\times\boldsymbol{k})\times\boldsymbol{e}_\mathrm{n}$ for the given $\boldsymbol{k}$ and $\boldsymbol{e}_\mathrm{n}$. Here the former is recommended due to the fact that the computational complexity of vector product is greater than that of scalar product.

## III. Relationship to Material Parameters

According to Maxwell's equations, the complex wave vector of a nonuniform plane wave propagating in a homogeneous, isotropic and linear medium satisfies the dispersion equation given by [29]

$$\boldsymbol{k}\cdot\boldsymbol{k} = k^2 = \omega^2\mu\varepsilon \tag{14}$$

where $k = \omega\sqrt{\mu\varepsilon}$ is the intrinsic wave number, $\varepsilon$ and $\mu$ are the permittivity and permeability of the medium, respectively.

For an ideally lossless medium, both of the two material parameters, $\varepsilon$ and $\mu$, are with real values, and therefore $k$ is also a real-valued wave number. However, for a lossy medium, if the time-harmonic dependency $\mathrm{e}^{-\mathrm{j}\omega t}$ is assumed, the complex permittivity $\varepsilon$ is given by

$$\varepsilon(\omega) = \varepsilon_0\varepsilon_\mathrm{r}(\omega) = \varepsilon'(\omega) + \mathrm{j}\varepsilon''(\omega) + \mathrm{j}\frac{\sigma}{\omega} \tag{15}$$

and the complex permeability $\mu$ is given by

$$\mu(\omega) = \mu_0\mu_\mathrm{r}(\omega) = \mu'(\omega) + \mathrm{j}\mu''(\omega) \tag{16}$$

where $\varepsilon_\mathrm{r}$ is the complex relative permittivity, $\mu_\mathrm{r}$ is the complex relative permeability, $\varepsilon''$ is the imaginary part of complex dielectric constant that represents polarization loss, $\sigma$ is the electrical conductivity that represents ohmic loss, and $\mu''$ is the imaginary part of complex permeability that represents magnetization loss. Therefore, $k$ becomes a complex-valued wave number for a lossy medium.

Substitution of (1) into (14) yields the constraint relationship among the phase constant $\beta$, the attenuation constant $\alpha$ and the angle $\varphi$ between $\boldsymbol{\beta}$ and $\boldsymbol{\alpha}$,

$$\omega^2\mu\varepsilon = k^2 = \boldsymbol{k}\cdot\boldsymbol{k} = (\boldsymbol{\beta}+\mathrm{j}\boldsymbol{\alpha})\cdot(\boldsymbol{\beta}+\mathrm{j}\boldsymbol{\alpha}) \\ = \beta^2-\alpha^2+2\mathrm{j}\boldsymbol{\beta}\cdot\boldsymbol{\alpha} = \beta^2-\alpha^2+2\mathrm{j}\beta\alpha\cos\varphi \quad (17)$$

with $\varphi = \arccos(\boldsymbol{e}_\beta\cdot\boldsymbol{e}_\alpha)$. Separating the real and imaginary parts of both sides of (17) yields the following two equations

$$\beta^2-\alpha^2 = \mathrm{Re}[k^2] = \omega^2\,\mathrm{Re}[\mu\varepsilon] \quad (18)$$

and

$$\beta\alpha\cos\varphi = \mathrm{Im}[k^2]/2 = \omega^2\,\mathrm{Im}[\mu\varepsilon]/2 \quad (19)$$

When the directions of $\boldsymbol{\beta}$ and $\boldsymbol{\alpha}$ of a nonuniform plane wave, or the angle $\varphi$ between them, are assigned, $\beta$ and $\alpha$ are determined by the set of solutions to (18) and (19) with real and positive values. However, for a lossless medium of $\mathrm{Im}[\mu\varepsilon]=0$, $\cos\varphi = 0$ i.e. $\varphi = \pi/2$. Thus the propagation vectors $\boldsymbol{\beta}$ and $\boldsymbol{\alpha}$ of a nonuniform plane wave in a lossless medium must be perpendicular to each other, such as the evanescent wave in total reflection at the interface of two lossless media.

On the other hand, from (11) we have

$$k^2 = k_{\mathrm{n}}^2 + k_{\mathrm{t}}^2 = (\beta_{\mathrm{n}}+\mathrm{j}\alpha_{\mathrm{n}})^2 + (\boldsymbol{\beta}_{\mathrm{t}}+\mathrm{j}\boldsymbol{\alpha}_{\mathrm{t}})\cdot(\boldsymbol{\beta}_{\mathrm{t}}+\mathrm{j}\boldsymbol{\alpha}_{\mathrm{t}}) \\ = \beta_{\mathrm{n}}^2-\alpha_{\mathrm{n}}^2+2\mathrm{j}\beta_{\mathrm{n}}\alpha_{\mathrm{n}}+\beta_{\mathrm{t}}^2-\alpha_{\mathrm{t}}^2+2\mathrm{j}\beta_{\mathrm{t}}\alpha_{\mathrm{t}}\cos\varphi_{\mathrm{t}} \\ = \beta^2-\alpha^2+2\mathrm{j}(\beta_{\mathrm{n}}\alpha_{\mathrm{n}}+\beta_{\mathrm{t}}\alpha_{\mathrm{t}}\cos\varphi_{\mathrm{t}}) \quad (20)$$

where $\varphi_{\mathrm{t}} = \arccos(\boldsymbol{e}_{\beta\mathrm{t}}\cdot\boldsymbol{e}_{\alpha\mathrm{t}})$ is the angle between $\boldsymbol{\alpha}_{\mathrm{t}}$ and $\boldsymbol{\beta}_{\mathrm{t}}$. By comparing (17) and (20), we obtain

$$\beta\alpha\cos\varphi = \beta_{\mathrm{n}}\alpha_{\mathrm{n}} + \beta_{\mathrm{t}}\alpha_{\mathrm{t}}\cos\varphi_{\mathrm{t}} \quad (21)$$

It is also seen from Fig. 1 that $\beta_{\mathrm{n}} = \beta\cos\theta_\beta$, $\alpha_{\mathrm{n}} = \alpha\cos\theta_\alpha$, $\beta_{\mathrm{t}} = \beta\sin\theta_\beta$ and $\alpha_{\mathrm{t}} = \alpha\sin\theta_\alpha$, so we finally get

$$\cos\varphi = \cos\theta_\beta\cos\theta_\alpha + \sin\theta_\beta\sin\theta_\alpha\cos\varphi_{\mathrm{t}} \quad (22)$$

This relationship conforms to the spherical trigonometry between $\boldsymbol{\beta}$ and $\boldsymbol{\alpha}$ given by (20) in Ref. [13]. Thus the validity of the decomposition of the complex wave vector and complex wave number proposed in Section II is verified.

## IV. Application to Snell's Law

To solve the reflection and transmission problems of a nonuniform plane wave incident on an arbitrarily oriented planar interface between two isotropic and possibly lossy media, the complex forms of Snell's law and Fresnel equations are utilized based on the complex angles of incidence and refraction as defined in Section II. For the sake of simplicity, the following notations are adopted: the physical quantities relevant to the incident wave are denoted without primes, the physical quantities relevant to the reflected wave are denoted by single primes, and the physical quantities relevant to the transmitted wave are denoted by double primes. Moreover, suppose that the incident wave impinges upon the interface from medium 1 to medium 2 and the unit normal vector $\boldsymbol{e}_{\mathrm{n}}$ of interface is also pointing from medium 1 to medium 2.

When the three vectors, $\boldsymbol{\beta}$, $\boldsymbol{\alpha}$ and $\boldsymbol{e}_{\mathrm{n}}$, are not coplanar as the case shown in Fig. 1, the incident nonuniform plane wave can be represented as the superposition of a parallel electric (PE) wave and a parallel magnetic (PM) wave. If the time-harmonic dependence $\mathrm{e}^{-\mathrm{j}\omega t}$ is suppressed, the complex electric fields of the incident, reflected and transmitted nonuniform plane waves are given by

$$\boldsymbol{E}(\boldsymbol{r}) = \boldsymbol{E}_0\,\mathrm{e}^{\mathrm{j}\boldsymbol{k}\cdot\boldsymbol{r}} = \boldsymbol{e}_{\mathrm{PE}}E_{\mathrm{PE}}(\boldsymbol{r}) + \boldsymbol{e}_{\mathrm{PM}}E_{\mathrm{PM}}(\boldsymbol{r}) \quad (23)$$

$$\boldsymbol{E}'(\boldsymbol{r}) = \boldsymbol{E}'_0\,\mathrm{e}^{\mathrm{j}\boldsymbol{k}'\cdot\boldsymbol{r}} = \boldsymbol{e}'_{\mathrm{PE}}E'_{\mathrm{PE}}(\boldsymbol{r}) + \boldsymbol{e}'_{\mathrm{PM}}E'_{\mathrm{PM}}(\boldsymbol{r}) \quad (24)$$

$$\boldsymbol{E}''(\boldsymbol{r}) = \boldsymbol{E}''_0\,\mathrm{e}^{\mathrm{j}\boldsymbol{k}''\cdot\boldsymbol{r}} = \boldsymbol{e}''_{\mathrm{PE}}E''_{\mathrm{PE}}(\boldsymbol{r}) + \boldsymbol{e}''_{\mathrm{PM}}E''_{\mathrm{PM}}(\boldsymbol{r}) \quad (25)$$

where $\boldsymbol{k} = \boldsymbol{\beta}+\mathrm{j}\boldsymbol{\alpha}$, $\boldsymbol{k}' = \boldsymbol{\beta}'+\mathrm{j}\boldsymbol{\alpha}'$ and $\boldsymbol{k}'' = \boldsymbol{\beta}''+\mathrm{j}\boldsymbol{\alpha}''$ are the complex wave vectors of the incident, reflected and transmitted waves, respectively. $\boldsymbol{E}_0$, $\boldsymbol{E}'_0$ and $\boldsymbol{E}''_0$ are the complex electric field at the original point $\boldsymbol{r}=0$, $E_{\mathrm{PE}}$, $E'_{\mathrm{PE}}$ and $E''_{\mathrm{PE}}$ are the complex electric field magnitudes of PE waves at an arbitrary point $\boldsymbol{r}$, while $E_{\mathrm{PM}}$, $E'_{\mathrm{PM}}$ and $E''_{\mathrm{PM}}$ are the complex electric fields of PM waves at an arbitrary point $\boldsymbol{r}$, $\boldsymbol{e}'_{\mathrm{PE}}$ and $\boldsymbol{e}''_{\mathrm{PE}}$ are the complex unit polarization vectors of the electric fields of PE waves given by

$$\boldsymbol{e}_{\mathrm{PE}} = \boldsymbol{e}'_{\mathrm{PE}} = \boldsymbol{e}''_{\mathrm{PE}} = \frac{\boldsymbol{s}}{\sqrt{\boldsymbol{s}\cdot\boldsymbol{s}}} \quad (26)$$

with $\boldsymbol{s} = \boldsymbol{e}_{\mathrm{n}}\times\boldsymbol{k}$ ,which are all parallel to the interface and "perpendicular" to the "complex plane of incidence" in the complex domain. $\boldsymbol{e}_{\mathrm{PM}}$, $\boldsymbol{e}'_{\mathrm{PM}}$ and $\boldsymbol{e}''_{\mathrm{PM}}$ are the complex unit polarization vectors of the electric fields of PM waves given by

$$\boldsymbol{e}_{\mathrm{PM}} = \frac{\boldsymbol{p}}{\sqrt{\boldsymbol{p}\cdot\boldsymbol{p}}},\ \boldsymbol{e}'_{\mathrm{PM}} = \frac{\boldsymbol{p}'}{\sqrt{\boldsymbol{p}'\cdot\boldsymbol{p}'}},\ \boldsymbol{e}''_{\mathrm{PM}} = \frac{\boldsymbol{p}''}{\sqrt{\boldsymbol{p}''\cdot\boldsymbol{p}''}} \quad (27)$$

with $\boldsymbol{p} = (\boldsymbol{e}_{\mathrm{n}}\times\boldsymbol{k})\times\boldsymbol{k}$, $\boldsymbol{p}' = (\boldsymbol{e}_{\mathrm{n}}\times\boldsymbol{k}')\times\boldsymbol{k}'$ and $\boldsymbol{p}'' = (\boldsymbol{e}_{\mathrm{n}}\times\boldsymbol{k}'')\times\boldsymbol{k}''$, which are in the "complex plane of incidence" and "perpendicular" to $\boldsymbol{k}$, $\boldsymbol{k}'$ and $\boldsymbol{k}''$, respectively. Note that $\boldsymbol{e}_{\mathrm{PE}}\cdot\boldsymbol{e}_{\mathrm{PM}} = 0$, $\boldsymbol{e}_{\mathrm{PE}}\cdot\boldsymbol{k} = 0$ and $\boldsymbol{e}_{\mathrm{PM}}\cdot\boldsymbol{k} = 0$ for the incident wave, so that $\boldsymbol{E}\cdot\boldsymbol{k} = 0$ and the two component fields $E_{\mathrm{PE}}$ and $E_{\mathrm{PM}}$ of electric fields are related to $\boldsymbol{E}$ by

$$E_{\mathrm{PE}}(\boldsymbol{r}) = \boldsymbol{e}_{\mathrm{PE}}\cdot\boldsymbol{E}(\boldsymbol{r}),\ E_{\mathrm{PM}}(\boldsymbol{r}) = \boldsymbol{e}_{\mathrm{PM}}\cdot\boldsymbol{E}(\boldsymbol{r}) \quad (28)$$

where $\boldsymbol{E}(\boldsymbol{r})$ is the complex electric field of the incident nonuniform plane wave at an arbitrary point $\boldsymbol{r}$. This fact also holds for the reflected and transmitted waves. Especially, when the three vectors, $\boldsymbol{\beta}$, $\boldsymbol{\alpha}$ and $\boldsymbol{e}_{\mathrm{n}}$, are coplanar, the PE and PM waves degenerate into the familiar TE and TM waves, respectively. In this situation, the unit vectors of the electric fields of the two polarization waves, $\boldsymbol{e}_{\mathrm{TE}}$ and $\boldsymbol{e}_{\mathrm{TM}}$, are with real values and have physically meaningful directions.

According to the boundary conditions of electromagnetic fields, the tangential components of the electric fields of the incident, reflected and transmitted waves must satisfy the phase-matching condition at any point on the interface,

$$\boldsymbol{k}\cdot\boldsymbol{r}=\boldsymbol{k}'\cdot\boldsymbol{r}=\boldsymbol{k}''\cdot\boldsymbol{r} \tag{29}$$

Suppose that $\boldsymbol{r}_1$ and $\boldsymbol{r}_2$ are any two points on the interface and by substituting them into (29), we have the relative position vector $\boldsymbol{r}_\text{s}=\boldsymbol{r}_2-\boldsymbol{r}_1$ that satisfies

$$\boldsymbol{k}\cdot\boldsymbol{r}_\text{s}=\boldsymbol{k}'\cdot\boldsymbol{r}_\text{s}=\boldsymbol{k}''\cdot\boldsymbol{r}_\text{s} \tag{30}$$

Note that the vector $\boldsymbol{r}_\text{s}$ is in the plane of interface, so that $\boldsymbol{e}_\text{n}\cdot\boldsymbol{r}_\text{s}=0$ and $\boldsymbol{k}\cdot\boldsymbol{r}_\text{s}=(\boldsymbol{k}_\text{t}+\boldsymbol{k}_\text{n})\cdot\boldsymbol{r}_\text{s}=(\boldsymbol{k}_\text{t}+k_\text{n}\boldsymbol{e}_\text{n})\cdot\boldsymbol{r}_\text{s}=\boldsymbol{k}_\text{t}\cdot\boldsymbol{r}_\text{s}$, as well as $\boldsymbol{k}'\cdot\boldsymbol{r}_\text{s}=\boldsymbol{k}'_\text{t}\cdot\boldsymbol{r}_\text{s}$ and $\boldsymbol{k}''\cdot\boldsymbol{r}_\text{s}=\boldsymbol{k}''_\text{t}\cdot\boldsymbol{r}_\text{s}$. Therefore based on (30), one obtains

$$\boldsymbol{k}_\text{t}\cdot\boldsymbol{r}_\text{s}=\boldsymbol{k}'_\text{t}\cdot\boldsymbol{r}_\text{s}=\boldsymbol{k}''_\text{t}\cdot\boldsymbol{r}_\text{s} \tag{31}$$

Since $\boldsymbol{k}_\text{t}$, $\boldsymbol{k}'_\text{t}$, $\boldsymbol{k}''_\text{t}$ and $\boldsymbol{r}_\text{s}$ are all in the same plane of interface and (31) holds true for any $\boldsymbol{r}_\text{s}$, the tangential component vectors of wave vectors must be equal, that is

$$\boldsymbol{k}_\text{t}=\boldsymbol{k}'_\text{t}=\boldsymbol{k}''_\text{t}=\boldsymbol{k}-(\boldsymbol{e}_\text{n}\cdot\boldsymbol{k})\boldsymbol{e}_\text{n} \tag{32}$$

Equivalently, based on (7), the tangential component vectors of the phase and attenuation vectors are also conserved that

$$\boldsymbol{\beta}_\text{t}=\boldsymbol{\beta}'_\text{t}=\boldsymbol{\beta}''_\text{t}=\boldsymbol{\beta}-(\boldsymbol{\beta}\cdot\boldsymbol{e}_\text{n})\boldsymbol{e}_\text{n} \tag{33}$$

and

$$\boldsymbol{\alpha}_\text{t}=\boldsymbol{\alpha}'_\text{t}=\boldsymbol{\alpha}''_\text{t}=\boldsymbol{\alpha}-(\boldsymbol{\alpha}\cdot\boldsymbol{e}_\text{n})\boldsymbol{e}_\text{n}. \tag{34}$$

Then according to (9), the tangential components of the complex wave numbers are also equal,

$$k_\text{t}=k'_\text{t}=k''_\text{t}=\sqrt{\boldsymbol{k}_\text{t}\cdot\boldsymbol{k}_\text{t}} \tag{35}$$

and the complex form of Snell's law is obtained,

$$k_1\sin\theta=k_1\sin\theta'=k_2\sin\theta''=k_\text{t} \tag{36}$$

where $k_1=\omega\sqrt{\mu_1\varepsilon_1}$ and $k_2=\omega\sqrt{\mu_2\varepsilon_2}$. Note that $k_1=k_0n_1$ and $k_2=k_0n_2$, the complex Snell's law also can be expressed in terms of refractive indices as

$$n_1\sin\theta=n_1\sin\theta'=n_2\sin\theta''=k_\text{t}/k_0 \tag{37}$$

where $k_0=\omega\sqrt{\mu_0\varepsilon_0}$ is the wave number in free space, $n_1=\sqrt{\mu_{1\text{r}}\varepsilon_{1\text{r}}}$ and $n_2=\sqrt{\mu_{2\text{r}}\varepsilon_{2\text{r}}}$ are the complex refractive indices of medium 1 and medium 2, respectively. Based on (36), the complex angles of incidence and reflection are defined as

$$\theta=\arcsin(k_\text{t}/k_1),\ \theta'=\pi-\theta \tag{38}$$

and the complex angle of refraction is defined as

$$\theta''=\arcsin(k_\text{t}/k_2) \tag{39}$$

Usually, the complex wave vector of the incident plane wave

$$\boldsymbol{k}=\boldsymbol{\beta}+\text{j}\boldsymbol{\alpha}=\beta\boldsymbol{e}_\beta+\text{j}\alpha\boldsymbol{e}_\alpha \tag{40}$$

is known at the first interface or can been calculated from the preceding interfaces. According to (12) and (38), $k'_\text{n}=k_1\cos\theta'=-k_1\cos\theta=-k_\text{n}$, and perforce the complex wave vector of the reflected wave is given by

$$\boldsymbol{k}'=\boldsymbol{k}'_\text{t}+\boldsymbol{k}'_\text{n}=\boldsymbol{k}_\text{t}-\boldsymbol{k}_\text{n}=\boldsymbol{k}_\text{t}-k_1\cos\theta\boldsymbol{e}_\text{n}. \tag{41}$$

Then one has $\boldsymbol{\beta}'=\text{Re}[\boldsymbol{k}']$ and $\boldsymbol{\alpha}'=\text{Im}[\boldsymbol{k}']$.

Similarly, substitution of $\theta''$ calculated by (39) into (12) yields $k''_\text{n}=k_2\cos\theta''$, so that the complex wave vector of the transmitted wave is obtained by

$$\boldsymbol{k}''=\boldsymbol{k}''_\text{t}+\boldsymbol{k}''_\text{n}=\boldsymbol{k}_\text{t}+k''_\text{n}\boldsymbol{e}_\text{n}=\boldsymbol{k}_\text{t}+k_2\cos\theta''\boldsymbol{e}_\text{n} \tag{42}$$

Then one has $\boldsymbol{\beta}''=\text{Re}[\boldsymbol{k}'']$ and $\boldsymbol{\alpha}''=\text{Im}[\boldsymbol{k}'']$.

## V. Fresnel Equations for Charged Interfaces

Now forward to the Fresnel equations for a nonuniform incident plane wave impinges on a charged planar interface. The boundary conditions of electromagnetic fields for the charged interface between two different media can be derived from the integral form of Maxwell's equations and listed as

$$\boldsymbol{e}_\text{n}\cdot[\varepsilon_2\boldsymbol{E}''-\varepsilon_1(\boldsymbol{E}+\boldsymbol{E}')]=\rho_\text{s} \tag{43}$$

$$\boldsymbol{e}_\text{n}\cdot[\mu_2\boldsymbol{H}''(\boldsymbol{r})-\mu_1(\boldsymbol{H}+\boldsymbol{H}')]=0 \tag{44}$$

$$\boldsymbol{e}_\text{n}\times[\boldsymbol{E}''-(\boldsymbol{E}+\boldsymbol{E}')]=0 \tag{45}$$

$$\boldsymbol{e}_\text{n}\times[\boldsymbol{H}''-(\boldsymbol{H}+\boldsymbol{H}')]=\boldsymbol{J}_\text{s} \tag{46}$$

where $\rho_\text{s}$ is the external (excess) surface charge density and $\boldsymbol{J}_\text{s}$ is the surface current density. Based on the complex angles of incidence and refraction, $\theta$ and $\theta''$, calculated by (38) and (39) and noting that $\boldsymbol{e}_\text{n}\cdot\boldsymbol{k}=k_\text{n}=k\cos\theta$ and $\boldsymbol{e}_\text{n}\cdot\boldsymbol{k}''=k''_\text{n}=k''\cos\theta''$, the Fresnel reflection and transmission coefficients for the incident nonuniform plane waves of PE polarization and PM polarization can be derived from (45) and (46) as [28]

$$\begin{cases} r_\text{PE}=\dfrac{E'_\text{PE}}{E_\text{PE}}=\dfrac{Z_2\cos\theta-Z_1\cos\theta''-\sigma_\text{s}Z_1Z_2}{Z_2\cos\theta+Z_1\cos\theta''+\sigma_\text{s}Z_1Z_2}\\ t_\text{PE}=\dfrac{E''_\text{PE}}{E_\text{PE}}=\dfrac{2Z_2\cos\theta}{Z_2\cos\theta+Z_1\cos\theta''+\sigma_\text{s}Z_1Z_2}\end{cases} \tag{47}$$

and

$$\begin{cases} r_\text{PM}=\dfrac{E'_\text{PM}}{E_\text{PM}}=\dfrac{Z_1/\cos\theta''-Z_2/\cos\theta+\sigma_\text{s}Z_1Z_2}{Z_1/\cos\theta''+Z_2/\cos\theta+\sigma_\text{s}Z_1Z_2}\\ t_\text{PM}=\dfrac{E''_\text{PM}}{E_\text{PM}}=\dfrac{2Z_2/\cos\theta''}{Z_1/\cos\theta''/+Z_2/\cos\theta+\sigma_\text{s}Z_1Z_2}\end{cases} \tag{48}$$

where $Z_1=\sqrt{\mu_1/\varepsilon_1}$ and $Z_2=\sqrt{\mu_2/\varepsilon_2}$ are the intrinsic impedances of the two media, respectively. $\sigma_\text{s}$ is the surface conductivity given by [30]

$$\sigma_s(\omega) = \frac{\rho_s q_s / m_s}{\gamma_s + \gamma_{rad}\omega^2 - j\omega} \quad (49)$$

with $\gamma_s = k_B T/\hbar$ and $\gamma_{rad} = q_s^2/(6\pi\varepsilon_0 m_s c^3)$, where $\rho_s$ is the external surface charge density, $q_s$ is the electric charge, $m_s$ is the mass of charge, $k_B$ is the Boltzmann constant, $T$ is the temperature in Kelvin, $\hbar$ is the reduced Planck constant and $c$ is the speed of light in free space. When a charged interface is assumed with $\sigma_s \neq 0$, the surface current density $\boldsymbol{J}_s = \sigma_s \boldsymbol{E}_{tan}$ stimulated by the tangential electric field $\boldsymbol{E}_{tan} = \boldsymbol{E}''_t$ at the interface alters the tangential boundary condition of magnetic fields and results in a different form of Fresnel formula. However, for an uncharged interface, the Fresnel equations reduce to the familiar form with $\sigma_s = 0$.

Based on the Fresnel equations given by (47) and (48), the complex amplitudes of the electric fields of PE polarization are calculated by $E'_{0PE} = r_{PE}E_{0PE}$ and $E''_{0PE} = t_{PE}E_{0PE}$, while the complex amplitudes of the electric fields of PM polarization are obtained by $E'_{0PM} = r_{PM}E_{0PM}$ and $E''_{0PM} = t_{PM}E_{0PM}$. Finally, the electric fields of the reflected and transmitted waves, $\boldsymbol{E}'(\boldsymbol{r})$ and $\boldsymbol{E}''(\boldsymbol{r})$, are obtained from (24) and (25) based on the calculated quantities. Meanwhile, the magnetic fields of the incident, reflected and transmitted waves can be calculated by

$$\boldsymbol{H} = \frac{\boldsymbol{k}\times\boldsymbol{E}}{\omega\mu_1},\ \boldsymbol{H}' = \frac{\boldsymbol{k}'\times\boldsymbol{E}'}{\omega\mu_1},\ \boldsymbol{H}'' = \frac{\boldsymbol{k}''\times\boldsymbol{E}''}{\omega\mu_2} \quad (50)$$

according to the Faraday's law of electromagnetic induction for plane waves, respectively.

## VI. Energy Flux and Energy Balance

Based on the complex electric and magnetic fields of the incident wave, reflected wave, and transmitted wave obtained above, we can further calculate the time-averaged energy flux densities represented by the time-averaged Poynting vectors. The time-averaged Poynting vectors in medium 1 is given by

$$\boldsymbol{S}_{av}^{M1} = \frac{1}{2}\mathrm{Re}[(\boldsymbol{E}+\boldsymbol{E}')\times(\boldsymbol{H}+\boldsymbol{H}')^*] = \boldsymbol{S}_{av} + \boldsymbol{S}'_{av} + \boldsymbol{S}_{av}^{mix} \quad (51)$$

where the asterisk mark (*) denotes the complex conjugate and

$$\boldsymbol{S}_{av} = \frac{1}{2}\mathrm{Re}[\boldsymbol{E}\times\boldsymbol{H}^*],\ \boldsymbol{S}'_{av} = \frac{1}{2}\mathrm{Re}[\boldsymbol{E}'\times\boldsymbol{H}'^*] \quad (52)$$

are the time-averaged Poynting vectors of the incident and reflected waves, respectively. $\boldsymbol{S}_{av}^{mix}$ is the mixed Poynting vector given by

$$\boldsymbol{S}_{av}^{mix} = \frac{1}{2}\mathrm{Re}[\boldsymbol{E}\times\boldsymbol{H}'^* + \boldsymbol{E}'\times\boldsymbol{H}^*] \quad (53)$$

which represents the interference of the incident and reflected waves in their overlapping region. In medium 2, there only exists the transmitted wave, so the time-averaged Poynting vectors is given by

$$\boldsymbol{S}_{av}^{M2} = \boldsymbol{S}''_{av} = \frac{1}{2}\mathrm{Re}[\boldsymbol{E}''\times\boldsymbol{H}''^*] \quad (54)$$

The surface Joule heat density at the interface contributed by the surface current is given by

$$p_s = \frac{1}{2}\mathrm{Re}[\boldsymbol{J}_s\cdot\boldsymbol{E}_t^*] = \frac{1}{2}\mathrm{Re}[\sigma_s\boldsymbol{E}_{tan}\cdot\boldsymbol{E}_{tan}^*] \quad (55)$$

where $\boldsymbol{E}_{tan} = \boldsymbol{E}''_t = \boldsymbol{E}'' - (\boldsymbol{e}_n\cdot\boldsymbol{E}'')\boldsymbol{e}_n$ is the tangential electric field at the interface and $\sigma_s$ is the surface conductivity.

Finally, the energy balance condition at the interface should be fulfilled by the relation given by

$$\boldsymbol{e}_n\cdot\boldsymbol{S}_{av}^{M1} = \boldsymbol{e}_n\cdot\boldsymbol{S}_{av}^{M2} + p_s \quad (56)$$

which is derived from the integral form of complex Poynting theorem by using a small Gaussian pillbox surrounding the interface. The correctness of all the above calculations can be verified by whether the two sides of (56) are equal.

## VII. Calculation Procedure

Based on the given parameters of incident plane wave, $\boldsymbol{E}_0$, $\boldsymbol{k}$ and $\omega$, the given material parameters of the two media, $\varepsilon'_{1,2}$, $\varepsilon''_{1,2}$, $\sigma_{1,2}$, $\mu'_{1,2}$ and $\mu''_{1,2}$, and the given parameters of interface, $\boldsymbol{e}_n$ and $\rho_s$, the corresponding reflected wave and transmitted wave can be calculated by the following steps:

1) Calculate the electromagnetic parameters $\varepsilon_{1,2}$ and $\mu_{1,2}$ according to (15) and (16) based on the given material parameters of the two media, then calculate the intrinsic wave numbers $k_{1,2}$ and the intrinsic impedances $Z_{1,2}$.
2) Based on the given complex wave vector $\boldsymbol{k} = \boldsymbol{\beta} + j\boldsymbol{\alpha}$ and the unit normal vector $\boldsymbol{e}_n$, calculate the tangent component vector $\boldsymbol{k}_t$ according to (32) and substitute it into (35) to calculate $k_t$. Further substitute $k_t$ into (38) and (39) to calculate the complex angles $\theta$, $\theta'$ and $\theta''$.
3) Calculate the complex wave vector $\boldsymbol{k}'$ of reflected wave based on (41), and then get $\boldsymbol{\beta}' = \mathrm{Re}[\boldsymbol{k}']$ and $\boldsymbol{\alpha}' = \mathrm{Im}[\boldsymbol{k}']$. Calculate the complex wave vector $\boldsymbol{k}''$ of transmitted wave based on (42), then get $\boldsymbol{\beta}'' = \mathrm{Re}[\boldsymbol{k}'']$ and $\boldsymbol{\alpha}'' = \mathrm{Im}[\boldsymbol{k}'']$.
4) Calculate the unit vectors $\boldsymbol{e}_{PE}$, $\boldsymbol{e}'_{PE}$ and $\boldsymbol{e}''_{PE}$ of the electric fields of PE waves according to (26). Calculate the unit vectors $\boldsymbol{e}_{PM}$, $\boldsymbol{e}'_{PM}$ and $\boldsymbol{e}''_{PM}$ of the electric fields of PM waves according to (27), respectively. Calculate the electric field magnitudes of the incident PE and PM waves, $E_{PE}$ and $E_{PM}$ by (28) from the given electric field $\boldsymbol{E}$.
5) Calculate the surface conductivity $\sigma_s$ based on the given external surface charge density $\rho_s$ according to (49). Calculate the reflection and transmission coefficients, $r_{PE}$ and $t_{PE}$ of PE waves according to (47) and therefore $E'_{PE} = r_{PE}E_{PE}$ and $E''_{PE} = t_{PE}E_{PE}$ are obtained. Calculate the reflection and transmission coefficients, $r_{PM}$ and $t_{PM}$ of PM waves according to (48) and therefore $E'_{\text{PM}} = r_{\text{PM}} E_{\text{PM}}$ and $E''_{\text{PM}} = t_{\text{PM}} E_{\text{PM}}$ are obtained.

6) Calculate the complex electric fields of the reflected and transmitted waves, $\boldsymbol{E}'$ and $\boldsymbol{E}''$ according to (24) and (25), respectively. Calculate the magnetic fields of the incident, reflected and transmitted waves by (50), respectively. Calculate the energy flux and Joule heat by (51)-(55), respectively, if needed.

## VIII. Examples and Results

A two-dimensional (2D) example of a charged and lossy prism placed in air when $\boldsymbol{\beta}''$, $\boldsymbol{\alpha}''$ and $\boldsymbol{e}_{\text{n}}$ are coplanar and a three-dimensional (3D) example of a cuboid composed of two lossy blocks when $\boldsymbol{\beta}''$, $\boldsymbol{\alpha}''$ and $\boldsymbol{e}_{\text{n}}$ are not coplanar are presented in this section to demonstrate the simplicity and strength of our proposed algorithm.

### *A. 2D Case when $\boldsymbol{\beta}''$, $\boldsymbol{\alpha}''$ and $\boldsymbol{e}_{\text{n}}$ are coplanar*

In the first example of 2D case where $\boldsymbol{\beta}''$, $\boldsymbol{\alpha}''$ and $\boldsymbol{e}_{\text{n}}$ are coplanar, we consider a circularly polarized uniform plane wave with frequency $f = 0.1$ GHz is obliquely incident on the left surface of a lossy and charged prism placed in air with an incident angle $\theta_1 = 45^\circ$, as shown in Fig. 2. The two electric field magnitudes of the TE and TM waves are $E_{0\text{TE}} = 1$ V/m and $E_{0\text{TM}} = 1\,\text{e}^{\text{j}\pi/2}$ V/m, respectively. The prism has a vertex angle $\phi = 30^\circ$ and is made of a nonmagnetic isotropic lossy medium with a complex refractive index $n = 2 + \text{j}0.25$ at frequency $f$. The transmitted wave at the first interface becomes a nonuniform plane wave and further impinges upon the second interface i.e. the hypotenuse side of the prism. The reference point on the first interface is $O_1$, based on which the Cartesian coordinates are established with the directions of the three axes depicted in Fig. 2, so $\boldsymbol{r}_1 = \boldsymbol{0}$. The reference point on the second interface is $O_2$, which is on the right of $O_1$ along the $x$ axis with a distance $L = 0.8$ m, so that $\boldsymbol{r}_2 = L\boldsymbol{e}_x$. The surface charge density of electrons at the two interfaces are assumed to be $\rho_{\text{s}1} = -1\ \text{C/m}^2$ and $\rho_{\text{s}2} = -0.98\ \text{C/m}^2$, respectively.

Based on the given material parameters, we have the intrinsic electromagnetic parameters,

$$\varepsilon_{1\text{r}} = \varepsilon_{3\text{r}} = 1,\ \varepsilon_{2\text{r}} = n_2^2,\ \mu_{1\text{r}} = \mu_{2\text{r}} = \mu_{3\text{r}} = 1$$

$$k_1 = k_3 = k_0,\ k_2 = k_0 n_2,\ Z_1 = Z_3 = Z_0,\ Z_2 = Z_0 / n_2$$

where $n_2 = 2 + \text{j}0.25$, $k_0 = 2\pi f\sqrt{\mu_0\varepsilon_0}$ and $Z_0 = \sqrt{\mu_0/\varepsilon_0}$. The unit normal vectors of the two interfaces and the surface conductivities at the two reference points, $O_1$ and $O_2$, at the room temperature $T = 20$ °C are

$$\boldsymbol{e}_{\text{n}1} = \boldsymbol{e}_x,\ \boldsymbol{e}_{\text{n}2} = \boldsymbol{e}_x \cos\phi + \boldsymbol{e}_z \sin\phi$$

$$\sigma_{\text{s}1} = 0.00522\ \text{S}\quad \sigma_{\text{s}2} = 0.00512\ \text{S}$$

with $\phi = 30^\circ$, respectively.

The relevant parameters of the incident plane wave at the reference point $O_1 = \boldsymbol{r}_1$ on the first interface are

$$\boldsymbol{e}_{k_1} = \cos\theta_1 \boldsymbol{e}_x + \sin\theta_1 \boldsymbol{e}_z,\ \boldsymbol{k}_1 = k_1 \boldsymbol{e}_{k_1}$$

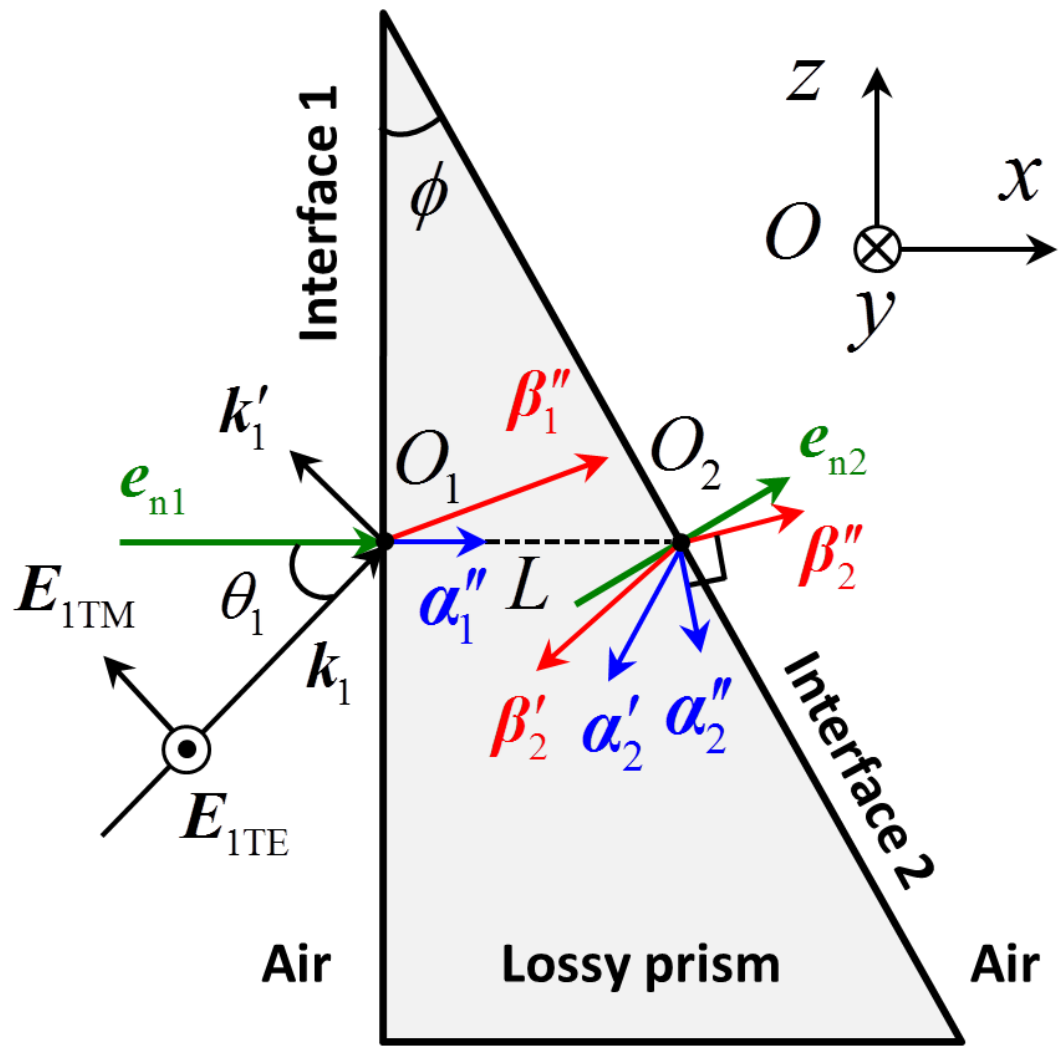

Fig. 2. A circularly polarized uniform plane obliquely impinges on a charged and lossy prism placed in air and the nonuniform plane waves are produced.

$$\boldsymbol{e}_{1\text{TE}} = -\boldsymbol{e}_y,\ \boldsymbol{e}_{1\text{TM}} = -\sin\theta_1 \boldsymbol{e}_x + \cos\theta_1 \boldsymbol{e}_z$$

$$\boldsymbol{E}_1 = \boldsymbol{E}_{1\text{TE}} + \boldsymbol{E}_{1\text{TM}} = \boldsymbol{e}_{1\text{TE}} + \boldsymbol{e}_{1\text{TM}}\,\text{e}^{\text{j}\pi/2}\ \text{V/m with } \theta_1 = 45^\circ.$$

Following the calculation procedure presented in Section VII, we obtain the results about the reflected and transmitted plane waves at the first interface, such as

$$\theta_1 = 45^\circ,\ \theta_1'' = 0.355 - \text{j}0.0464\ \text{rad},\ \theta''_{\alpha 1} = 0^\circ,\ \theta''_{\beta 1} = 20.7^\circ$$

$$\boldsymbol{k}_1' = \boldsymbol{\beta}_1' = -1.48\boldsymbol{e}_x + 1.48\boldsymbol{e}_z\ \text{rad/m}$$

$$\boldsymbol{k}_1'' = \boldsymbol{\beta}_1'' + \text{j}\boldsymbol{\alpha}_1'' = (3.93\boldsymbol{e}_x + 1.48\boldsymbol{e}_z) + \text{j}0.559\boldsymbol{e}_x\ \text{rad/m}$$

$$\boldsymbol{E}_1' = 0.345\text{e}^{\text{j}-1.53}\boldsymbol{e}_x + 0.690\text{e}^{\text{j}0.0264}\boldsymbol{e}_y + 0.345\text{e}^{\text{j}-1.53}\boldsymbol{e}_z\ \text{V/m}$$

$$\boldsymbol{E}_1'' = 0.135\text{e}^{-\text{j}1.75}\boldsymbol{e}_x + 0.312\text{e}^{\text{j}3.08}\boldsymbol{e}_y + 0.362\text{e}^{\text{j}1.53}\boldsymbol{e}_z\ \text{V/m}$$

According to the parameters of the refracted wave at the first interface, we can calculate the electric field of the incident wave $\boldsymbol{E}_2$ at the reference point $O_2 = \boldsymbol{r}_2$ on the second interface, where $\boldsymbol{k}_2 = \boldsymbol{k}_1''$ and

$$\boldsymbol{E}_2 = \boldsymbol{E}_1''\text{e}^{\text{j}\boldsymbol{k}_2\cdot(\boldsymbol{r}_2 - \boldsymbol{r}_1)} = \boldsymbol{E}_1''\text{e}^{\text{j}\boldsymbol{k}_2^*\cdot(L\boldsymbol{e}_x)} = \boldsymbol{E}_{2\text{TE}} + \boldsymbol{E}_{2\text{TM}}$$
$$= 0.0865\text{e}^{\text{j}1.39}\boldsymbol{e}_x + 0.198\text{e}^{-\text{j}0.0595}\boldsymbol{e}_y + 0.232\text{e}^{-\text{j}1.61}\boldsymbol{e}_z\,\text{V/m}$$

Applying the calculation procedure once again, we finally obtain the parameters of the reflected wave, the transmitted plane wave, the energy flux density and the surface Joule heat density at the second interface, such as

$$\theta_2 = 0.168 + \text{j}0.0464,\ \theta_2'' = 0.327 + \text{j}0.140$$

$$\theta_{\beta 2} = 9.32^\circ,\ \theta'_{\beta 2} = 170.7^\circ,\ \theta''_{\beta 2} = 18.73^\circ\quad \theta_{\alpha 2} = 30^\circ,$$

$\theta'_{\alpha 2} = 150^\circ$ , $\theta''_{\alpha 2} = 108.7^\circ$

$\boldsymbol{k}'_2 = (-3.25\boldsymbol{e}_x - 2.66\boldsymbol{e}_z) + \mathrm{j}(-0.280\boldsymbol{e}_x - 0.484\boldsymbol{e}_z)$ rad/m

$\boldsymbol{k}''_2 = (2.08\boldsymbol{e}_x + 0.414\boldsymbol{e}_z) + \mathrm{j}(0.0577\boldsymbol{e}_x - 0.2906\boldsymbol{e}_z)$ rad/m

$\boldsymbol{E}'_2 = 0.0307\mathrm{e}^{-\mathrm{j}1.87}\boldsymbol{e}_x + 0.0388\mathrm{e}^{\mathrm{j}2.75}\boldsymbol{e}_y + 0.0367\mathrm{e}^{\mathrm{j}1.18}\boldsymbol{e}_z$ V/m

$\boldsymbol{E}''_2 = 0.0500\mathrm{e}^{\mathrm{j}1.01}\boldsymbol{e}_x + 0.162\mathrm{e}^{\mathrm{j}0.0187}\boldsymbol{e}_y + 0.206\mathrm{e}^{-\mathrm{j}1.49}\boldsymbol{e}_z$ V/m

$\boldsymbol{S}_{\mathrm{av}}^{\mathrm{M2}} = \left(2.48\boldsymbol{e}_x - 0.109\boldsymbol{e}_y + 0.842\boldsymbol{e}_z\right)\times 10^{-4}$ W/m$^2$

$\boldsymbol{S}_{\mathrm{av}}^{\mathrm{M3}} = \boldsymbol{S}''_{\mathrm{av}2} = \left(9.12\boldsymbol{e}_x - 1.27\boldsymbol{e}_y + 1.82\boldsymbol{e}_z\right)\times 10^{-5}$ W/m$^2$

$p_{\mathrm{s}2} = 1.68\times 10^{-4}$ W/m$^2$

where the numerical values are retained with 3 significant digits. It can be seen that $\theta''_{\alpha 2} - \theta''_{\beta 2} = 90^\circ$ , which conforms to the property of a nonuniform plane wave in a lossless medium that $\boldsymbol{\alpha}''_2$ must be perpendicular to $\boldsymbol{\beta}''_2$ , as depicted in Fig. 2. It also can be verified that the above results satisfy the energy balance given by (56) at the second interface.

### B. 3D Case when $\boldsymbol{\beta}''$ , $\boldsymbol{\alpha}''$ and $\boldsymbol{e}_{\mathrm{n}}$ are not coplanar

Now consider the most complex 3D case with an arbitrary nonuniform plane wave impinges on an arbitrarily oriented and charged planar interface between two lossy media where the phase vector $\boldsymbol{\beta}$, the attenuation vector $\boldsymbol{\alpha}$ and the unit normal vector $\boldsymbol{e}_{\mathrm{n}}$ are not coplanar. As illustrated in Fig. 3, a uniform plane wave with an arbitrary propagation direction is incident on a cuboid formed by the combination of two charged blocks made of two different lossy media. The transmitted nonuniform plane wave produced from the first interface with the unit normal vector $\boldsymbol{e}_{\mathrm{n}1} = \boldsymbol{e}_z$ further impinges on the second interface, which passes through the body diagonal of the cuboid and with an arbitrarily oriented unit normal vector $\boldsymbol{e}_{\mathrm{n}2}$ when the length, width, and height of the cuboid are elaborately assigned. Finally, the nonuniform plane wave transmitted from the third interface with the unit normal vector $\boldsymbol{e}_{\mathrm{n}3} = \boldsymbol{e}_z$ is obtained.

In the following, the unit direction vector of an arbitrarily oriented vector $\boldsymbol{A}$ in the 3-D space is expressed as

$$\boldsymbol{e}_A = \sin\theta_A\cos\varphi_A\boldsymbol{e}_x + \sin\theta_A\sin\varphi_A\boldsymbol{e}_y + \cos\theta_A\boldsymbol{e}_z$$

where $\theta_A$ is the polar angle and $\varphi_A$ is the azimuth angle as those defined in the spherical coordinate system. Suppose that the incident uniform plane wave with frequency $f = 1$ THz is elliptically polarized with the two complex electric field magnitudes $E_{0\mathrm{PE}} = 1\,\mathrm{e}^{\mathrm{j}\pi/3}$ V/m and $E_{0\mathrm{PM}} = 2\,\mathrm{e}^{\mathrm{j}\pi/6}$ V/m of the PE and PM polarizations at the first point $O_1$ located on the center of the top surface, where the Cartesian coordinates are established and shown in Fig. 3. The polar and azimuth angles of the wave vector $\boldsymbol{k}_1$ of the incident plane wave are assumed to be $\theta_1 = 20^\circ$ and $\varphi_1 = -30^\circ$ , respectively. So that we have the parameters of the incident wave,

$$\boldsymbol{e}_{k_1} = \sin\theta_1\cos\varphi_1\boldsymbol{e}_x + \sin\theta_1\sin\varphi_1\boldsymbol{e}_y + \cos\theta_1\boldsymbol{e}_z$$

$$\boldsymbol{k}_1 = 2\pi f\sqrt{\mu_0\varepsilon_0}\,\boldsymbol{e}_{k_1} = (0.621\boldsymbol{e}_x - 0.358\boldsymbol{e}_y + 1.97\boldsymbol{e}_z)\times 10^4 \text{ rad/m}$$

$$\boldsymbol{e}_{1\mathrm{PE}} = 0.5\boldsymbol{e}_x + 0.866\boldsymbol{e}_y,\ \boldsymbol{e}_{1\mathrm{PM}} = 0.814\boldsymbol{e}_x - 0.470\boldsymbol{e}_y - 0.342\boldsymbol{e}_z$$

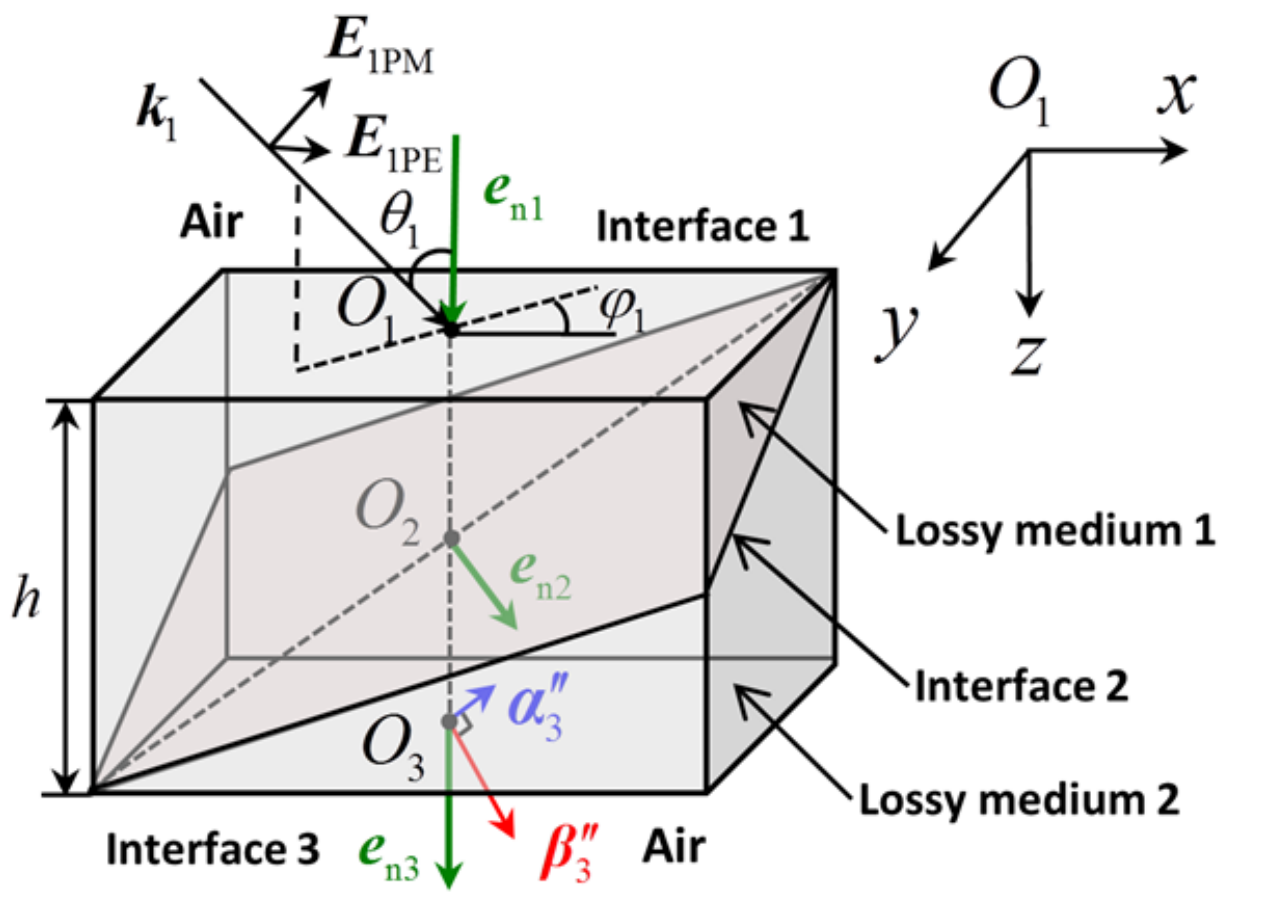

Fig. 3. An elliptically polarized uniform plane wave with an arbitrary propagation direction impinges on a cuboid formed by the combination of two lossy and charged blocs where the second interface is arbitrarily oriented.

$$\boldsymbol{E}_1 = \boldsymbol{E}_{1\mathrm{PE}} + \boldsymbol{E}_{1\mathrm{PM}} = 1\,\mathrm{e}^{\mathrm{j}\pi/3}\,\boldsymbol{e}_{1\mathrm{PE}} + 2\,\mathrm{e}^{\mathrm{j}\pi/6}\,\boldsymbol{e}_{1\mathrm{PM}} \text{ V/m}$$

If the height of the cuboid is $h = 20$ μm , the position vectors of the three reference points, $O_1$ , $O_2$ and $O_3$ , are

$$\boldsymbol{r}_1 = 0,\ \boldsymbol{r}_2 = h/2\boldsymbol{e}_z,\ \boldsymbol{r}_3 = h\boldsymbol{e}_z$$

respectively. The material parameters of the two lossy media at frequency $f$ are arbitrarily assigned with the values

$\varepsilon'_{\mathrm{r}1} = 2$, $\varepsilon''_{\mathrm{r}1} = 0.1$, $\sigma_1 = 0.2$ S/m, $\mu'_{\mathrm{r}1} = 1.2$, $\mu''_{\mathrm{r}1} = 0.3$ and

$\varepsilon'_{\mathrm{r}2} = 2.25$, $\varepsilon''_{\mathrm{r}2} = 0.4$, $\sigma_2 = 0.5$ S/m, $\mu'_{\mathrm{r}2} = 1.5$, $\mu''_{\mathrm{r}2} = 0.6$

The surface conductivities of the three interfaces and the unit normal vector of the second interface are assumed to be

$$\sigma_{\mathrm{s}1} = \sigma_{\mathrm{s}3} = (1 + 0.2\mathrm{j})\times 10^{-3}\text{ S},\ \sigma_{\mathrm{s}2} = (5 + \mathrm{j})\times 10^{-3}\text{ S}$$

and

$$\boldsymbol{e}_{\mathrm{n}2} = \sin\theta_{\mathrm{n}2}\cos\varphi_{\mathrm{n}2}\boldsymbol{e}_x + \sin\theta_{\mathrm{n}2}\sin\varphi_{\mathrm{n}2}\boldsymbol{e}_y + \cos\theta_{\mathrm{n}2}\boldsymbol{e}_z$$

with $\theta_{\mathrm{n}2} = 39^\circ$ and $\varphi_{\mathrm{n}2} = -21^\circ$ , respectively.

Applying the calculation procedures on the three interfaces in sequence, we get the calculation results about the transmitted plane wave, the energy flux density and the surface Joule heat density at the third interface, such that

$\theta''_3 = 0.553 + \mathrm{j}0.239$ rad , $\boldsymbol{r}_3 = h\boldsymbol{e}_z = 2\times 10^{-5}\boldsymbol{e}_z$ m

$\theta_{\beta''_3} = 31.7^\circ$ , $\varphi_{\beta''_3} = -26.7^\circ$ , $\theta_{\alpha''_3} = 122^\circ$ , $\varphi_{\alpha''_3} = -21^\circ$

$$\boldsymbol{k}''_3 = \boldsymbol{\beta}''_3 + \mathrm{j}\boldsymbol{\alpha}''_3 = (1.01\boldsymbol{e}_x - 0.509\boldsymbol{e}_y + 1.83\boldsymbol{e}_z)\times 10^4 + \mathrm{j}(0.403\boldsymbol{e}_x - 0.155\boldsymbol{e}_y - 0.266\boldsymbol{e}_z)\times 10^4 \text{ rad/m}$$

$$\boldsymbol{E}''_3 = 0.727\mathrm{e}^{\mathrm{j}1.08}\boldsymbol{e}_x + 0.186\mathrm{e}^{\mathrm{j}3.02}\boldsymbol{e}_y + 0.446\mathrm{e}^{-\mathrm{j}1.66}\boldsymbol{e}_z \text{ V/m}$$

$$\boldsymbol{H}''_3 = (0.400\mathrm{e}^{\mathrm{j}0.460}\boldsymbol{e}_x + 2.13\mathrm{e}^{\mathrm{j}1.17}\boldsymbol{e}_y + 0.443\mathrm{e}^{\mathrm{j}1.92}\boldsymbol{e}_z)\times 10^{-3} \text{ A/m}$$

$$\boldsymbol{S}_{\rm av}^{\rm M3} = \left(3.74\boldsymbol{e}_x - 1.31\boldsymbol{e}_y + 10.8\boldsymbol{e}_z\right)\times 10^{-4}\ \mathrm{W/m^2}$$

$$\boldsymbol{S}_{\rm av}^{\rm M4} = \boldsymbol{S}''_{\rm av3} = \left(4.70\boldsymbol{e}_x - 1.53\boldsymbol{e}_y + 8.03\boldsymbol{e}_z\right)\times 10^{-4}\ \mathrm{W/m^2}$$

$$p_{\rm s3} = 2.84\times 10^{-4}\ \mathrm{W/m^2}$$

where the numerical values are also retained with 3 significant digits. It can be seen that the relations $\boldsymbol{k}''_3\cdot\boldsymbol{E}''_3 = 0$, $\boldsymbol{\beta}''_3\cdot\boldsymbol{\alpha}''_3 = 0$, and the energy balance equation (56) are all satisfied, which verify the validity of the obtained results.

## IX. Conclusion

In this work, the decomposition of the complex wave vector and complex wave number of a nonuniform plane wave with respect to an arbitrarily oriented plane of interface is performed. The novel formula for calculating the tangential component of complex wave number provides the clear understanding and exact definitions of the complex angles of incidence, reflection and refraction with respect to the unit normal vector of the planar interface. This manipulation based on the complex vector analysis greatly simplifies the calculation of complex wave vectors and complex angles of the reflected and refracted waves for a given incident wave and facilitate their applications in the complex Snell's law and Fresnel formula. Compared to the various methods reported in literature, our method is the simplest one, especially in solving the most challenging problem when the phase vector $\boldsymbol{\beta}$, the attenuation vector $\boldsymbol{\alpha}$ and the unit normal vector $\boldsymbol{e}_{\rm n}$ are not coplanar. The given formulas and calculation procedure are very useful and powerful for analyzing the reflection and refraction phenomena of a nonuniform plane wave, including the evanescent wave generated by total reflection, incident on an arbitrarily oriented and charged plane boundary between two isotropic and possibly lossy media. This work also greatly deepened the understanding of the complex angles in the Snell's law and Fresnel's formulas. In fact, the proposed method can be applied to solve the reflection and refraction problems of any type of plane waves at the possibly charged interface between any types of isotropic and possibly lossy medium. This work may find applications in the fields of leaky-wave antenna, remote sensing, light ray-tracing software and the interaction of electromagnetic waves with biological tissues or metasurfaces.

## References

[1] F. Frezza and N. Tedeschi, "Electromagnetic inhomogeneous waves at planar boundaries: tutorial," *J. Opt. Soc. Amer. A*, vol. 32, no. 8, pp. 1485-1501, Aug. 2015.

[2] Y. Wang, A. Helmy, and G. Eleftheriades, "Ultra-wideband optical leaky wave slot antennas," *Opt. Exp.*, vol. 19, no. 13, pp. 12392-12401, 2011.

[3] R. Brem and T. F. Eibert, "A shooting and bouncing ray (SBR) modeling framework involving dielectrics and perfect conductors," *IEEE Trans. Antennas Propag.*, vol. 63, no. 8, pp. 3599-3609, Aug. 2015.

[4] Z. Cong, Z. He, and R. Chen, "An efficient volumetric SBR method for electromagnetic scattering from in-homogeneous plasma sheath," *IEEE Access*, vol. 7, pp. 90162-90170, Jul. 2019.

[5] Y. Huang, Z. Zhao, X. Li, Z. Nie, and Q.-H. Liu, "Volume equivalent SBR method for electromagnetic scattering of dielectric and composite objects," *IEEE Trans. Antennas Propag.*, vol. 69, no. 5, pp. 2842-2852, May 2021.

[6] A. Calcaterra, P. Simeoni , M. D. Migliore , F. Mangini, and F. Frezza, "Optimized leaky-wave antenna for hyperthermia in biological tissue theoretical model," *Sensors*, vol. 23, no. 21, Nov. 2023, Art. no. 8923.

[7] X. M. Mitsalas, T. Kaifas, and G. A. Kyriacou, "Space and leaky wave radiation from highly lossy biological cylindrical human-limps models," *Prog. Electromagn. Res. B*, vol. 94, pp. 145-174, Dec. 2021.

[8] S. Perea-Puente and F. J. Rodríguez-Fortuño, "Complex wave-vectors in lossy materials: From polarisation-loss locking to bullseye metasurface," *Proc. SPIE*, vol. 12131, pp. 63-72, May 2022.

[9] N. Tedeschi, V. Pascale, F. Pelorossi, and F. Frezza, "Generation of inhomogeneous electromagnetic waves by a lossy prism," *2016 URSI International Symposium on Electromagnetic Theory (EMTS)*, Espoo, Finland, 2016, pp. 838-841.

[10] R. B. Adler, L. J. Chu, and R. M. Fano, *Electromagnetic Energy Transmission and Radiation.* New York: Wiley, 1960, ch. 8.

[11] J. J. Holmes and C. A. Balanis, "Refraction of a uniform plane wave incident on a plane boundary between two lossy media," *IEEE Trans. Antennas Propagat.*, vol. AP-26, no.5, pp. 738-741, Sep. 1978.

[12] R. D. Radcliff and C. A. Balanis, "Modified propagation constants for nonuniform plane wave transmission through conducting media," *IEEE Trans. Geosci. Remote Sensing*, vol. GE-20, no.3, pp. 408-411, Jul. 1982.

[13] J. E. Roy, "New results for the effective propagation constants of nonuniform plane wave at the planar interface of two lossy media," *IEEE Trans. Antennas Propag.*, vol. 51, no. 6, pp. 1206-1215, Jun. 2003.

[14] J. E. Roy, "Corrections to 'New results for the effective propagation constants of nonuniform plane wave at the planar interface of two lossy media'," *IEEE Trans. Antennas Propag.*, vol. 53, no. 10, pp. 3420, Oct. 2005.

[15] F. Frezza and N. Tedeschi, "On the electromagnetic power transmission between two lossy media: Discussion," *J. Opt. Soc. Am. A*, vol. 29, no. 11, pp. 2281-2288, Nov. 2012.

[16] Fabio Mangini and Fabrizio Frezza, "Analysis of the electromagnetic reflection and transmission through a stratified lossy medium of an elliptically polarized plane wave," *Math. Mech. Complex Sy.*, vol. 4, no. 2, pp. 153-167, Nov. 2016.

[17] F. Frezza and N. Tedeschi, "Deeply penetrating waves in lossy media," *Opt. Lett.*, vol. 37, no. 13, pp. 2616–2618, Jul. 2012.

[18] P. Baccarelli, F. Frezza, P. Simeoni, and N. Tedeschi, "An analytical study of electromagnetic deep penetration conditions and implications in lossy media through inhomogeneous waves," *Materials*, vol. 11, no. 9, p. 1595, Sep. 2018.

[19] P. Baccarelli *et al.*, "Verification of the electromagnetic deep-penetration effect in the real world," *Sci. Rep.*, vol. 11, no. 1, Aug. 2021, Art. no. 15928.

[20] Y. Kim, H. Yang, and J. Oh, "Critical angle formulation of nonuniform plane waves for determining correct refraction angles at planar interface," *IEEE Trans. Antennas Propagat.*, vol. 71, no. 3, pp. 2861-2866, Mar. 2023.

[21] J. D. Jackson, *Classical Electrodynamics,* 3rd ed. New York: Wiley, 1999, Ch. 7.

[22] M. A. Dupertuis, M. Proctor, and B. Acklin, "Generalization of complex Snell-Descartes and Fresnel laws," *J. Opt. Soc. Am. A*, vol. 11, no. 3, pp. 1159-1166, Mar. 1994.

[23] V. Y. Fedorov and T. Nakajima, "Negative refraction of inhomogeneous waves in lossy isotropic media," *J. Opt.* vol. 16, Feb. 2014, Art. no. 035103.

[24] F. X. Canning, "Corrected Fresnel coefficients for lossy materials," *IEEE International Symposium on Antennas and Propagation,* Spokane, WA, 2011, pp. 2133-2136.

[25] I. M. Besieris, "Comment on the 'Corrected Fresnel coefficients for lossy materials'," *IEEE Antennas Propag. Mag.*, vol. 53, no. 4, pp. 161-164, Aug. 2011.

[26] H. Weber, "The Fresnel equations for lossy dielectrics and conservation of energy," *J. Modern Opt.,* vol. 61, no. 15, pp. 1219-1224, Jun. 2014

[27] M. C. Schake, "Comparison of recent results for the determination of effective propagation constants at attenuating interfaces," *Eur. J. Phys.*, vol. 39, Dec. 2017, Art. no. 015302.

[28] S. Zhang, L. Liu, and Y. Liu, "Generalized laws of Snell, Fresnel and energy balance for a charged planar interface between lossy media," *J. Quant. Spectrosc. Radiat. Transf.,* vol. 245, 2020, Art. no. 106903.

[29] J. -M. Jin, *Theory and Computation of Electromagnetic Fields*, 2nd ed. Hoboken, NJ: John Wiley & Sons, 2015, pp. 144-147.

[30] J. Klačka, M. Kocifaj, F. Kundracik, and G. Videen, "Optical signatures of electrically charged particles: Fundamental problems and solutions," *J. Quant. Spectrosc. Radiat. Transf.*, vol. 164, pp. 45-53, Oct. 2015.

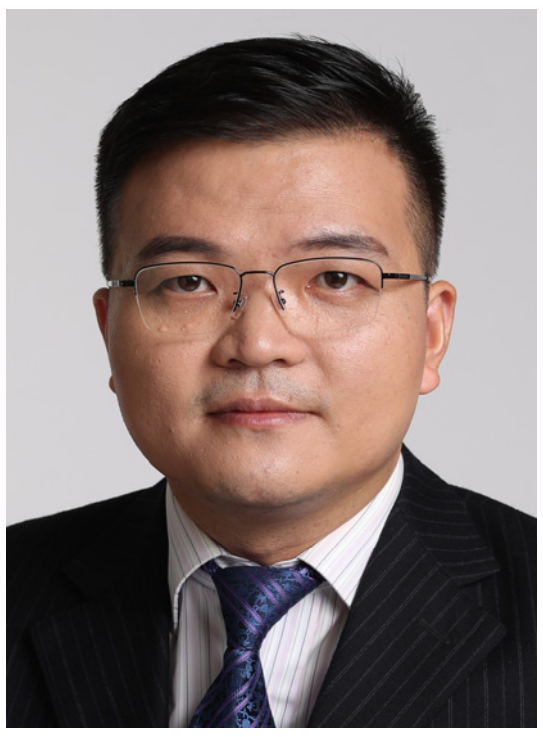

**Zhili Lin** (M'10–SM'12) received the B.S. degree in Optoelectronic Information Engineering in 2002 and the Ph.D. degree in Optical Engineering from Zhejiang University, Hangzhou, China in 2007.

From 2007 to 2008, he was a Postdoctoral Research Fellow with the Royal Institute of Technology (KTH), Sweden. From 2009 to 2013, he was a faculty member of the School of Instrumentation Science and Optoelectronics Engineering, Beihang University, Beijing, China. He is currently a Professor with the College of Information Science and Engineering, Huaqiao University, Xiamen, China. His research interests include computational electromagnetics and physical optics. He has authored or coauthored more than 120 papers in refereed journals.

Prof. Lin is a Senior Member of the Institute of Electrical and Electronics Engineers (IEEE), the Optical Society of America (OPTICA), the International Society for Optical Engineering (SPIE) and the Chinese Optical Society (COS).